\documentclass[a4paper,11pt,twocolumn]{article}
\usepackage{natbib}\usepackage{txfonts}\usepackage{balance}
\usepackage{graphicx}
\usepackage[a4paper]{hyperref}
\begin{document}
\title{FrogEye, the Quantum Coronagraphic mask:\\ \small{The Photon
Orbital Angular Momentum and its applications to Astronomy}}

\author{F. Tamburini, G. Umbriaco, G. Anzolin, C. Barbieri
and A. Bianchini \\ Department of Astronomy, University of Padova,
vicolo dell' Osservatorio 2, Padova, Italy}
\maketitle

\begin{abstract}
We propose to realize an  optical device based on the
properties of photon orbital angular momentum (POAM) to detect the
presence of closeby faint companions in double systems using
Laguerre-Gaussian (L-G) modes of the light. We test also the
possibility of using L-G modes to build coronagraph mask.  We
realized in the laboratory a prototype using a blazed l=1 hologram
to simulate the separation between two stars, as observed with a
telescope, in Laguerre-Gaussian modes.

keywords: Instrumentation: miscellaneous, Methods: laboratory,
Techniques: miscellaneous

\end{abstract}

\section{Introduction}

Photon orbital angular momentum (POAM) [\cite{al92,al00}] has
attracted attention for astronomical applications [\cite{har03}]
and applications at the quantum level for quantum communication
[\cite{mai01,vaz02,arl98}]. The generation of beams carrying POAM,
described by Laguerre-Gaussian (L-G) modes, proceeds thanks to the
insertion in the optical path of a phase modifying device (fork
hologram) imprinting a certain vorticity on the phase distribution
of the incident beam [\cite{arl99} ]. The vorticity is
characterized by parameters $l$ and $p$ describing the orbital and
azimuthal angular momentum. Here we describe a prototype of a
finger mask for binary systems using a $l=1$ fork-hologram.

\section{The Coronagraph}

The coronagraphic mask setup was tested using as source two 632nm
He-Ne lasers collimated into our $l=1$ fork--hologram simulating a
binary system having for each star angular diameter of 20marcsec,
namely the one produced by a 8.2m VLT-like telescope with adaptive
optics. The primary star lies on the center of the optical system,
on axis and crossing the fork-hologram singularity, while the
light beam representing the secondary star is off-axis. Tilted and
displaced beams do not possess a single L-G mode, but show a
spectrum of modes that depend on their geometrical properties
[\cite{vat}], which makes possible the detection of a fainter
companion.
\begin{figure*}[btc]
\resizebox{15cm}{!}{\includegraphics[width=15cm]{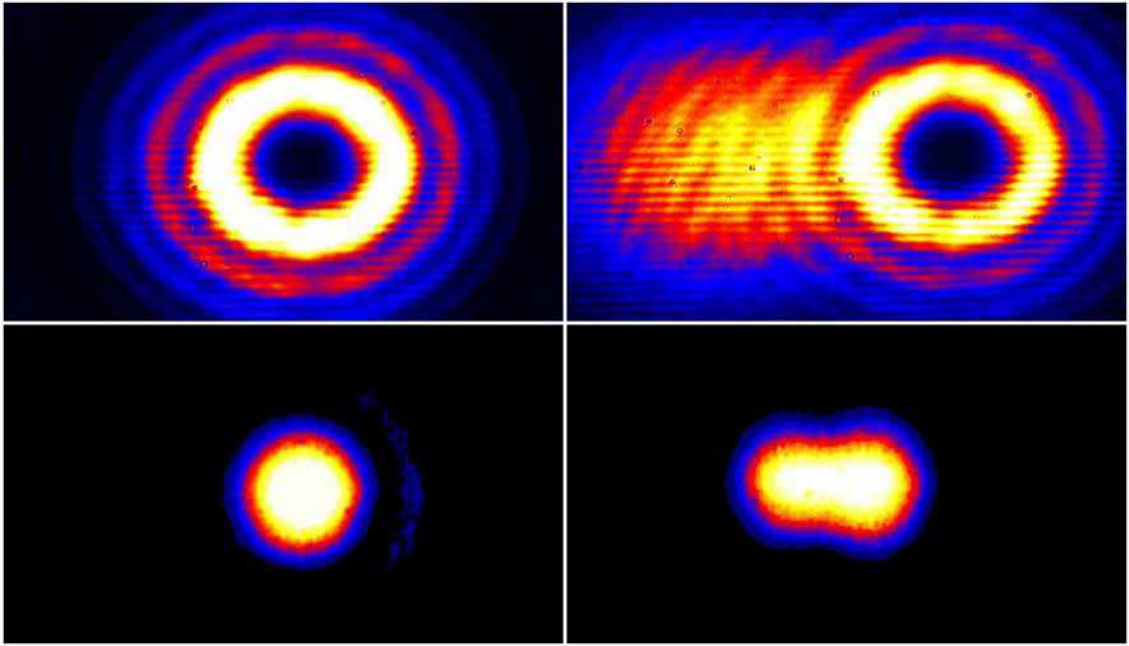}}
\caption{\footnotesize Upper row, left panel, L-G modes of an
unresolved binary composed of two equal intensity stars, right
panel, the same pair just resolved according to the Rayleigh
criterium. Bottom row: the corresponding gaussian modes. }
\label{lg1}
\end{figure*}

The light of the main star is spread over the donought-shaped
disk, while the fainter companion appears as a modification of the
L-G mode structures of the primary star. By setting the stars at a
very small distance for which the star is unresolved, we note that
the L-G profile of both the contributions has its characteristic
donought-symmetric shape. When the fainter star is resolved
following the Rayleigh limit, we see that the first outer ring
surrounding the donought becomes brighter and the level of the
background in the ring region rises up the more the separated are
the sources, thus revealing the presence of a fainter companion. A
fork-shaped hologram is equivalent to a set of cylindric lenses as
currently used in modern coronagraphs for extrasolar planets and
have the same optical properties of modifying the phase of the
light. Further tests are currently being performed to improve the
contrast of our finger-mask.

\section{Conclusions}
Photon distribution of light having non null orbital angular
momentum presents certain features that could give some advantages
to astronomical applications, such as giving better evidence to
closeby or fainter sources in a double system.

\section{acknowledgements}
The authors would like to acknowledge the support of ESO -
feasibility study of OWL instruments and Zeilinger Group,
University of Vienna, for the helpful discussions
and support for this work.

\bibliographystyle{aa}

\begin{thebibliography}{}

\bibitem[Allen et al. (1992)]{al92}
Allen, L. and Beijersbergen, M. W. and Spreew, R. J. C. and
Woerdman, J. P., PR, 45, 11, 8185-8189, 1992

\bibitem[Allen et al. (2002)]{al00} Allen, L. and Padgett,
M. J., OC, 184, 67-71, 2000

\bibitem[Harwit (2003)]{har03}  Harwit, M.,
ApJ, 597, 1266-1270, 2003

\bibitem[Arlt et al. (1998)]{arl98}
Arlt, J., Dholakia, K., Allen, L. and Padgett, M. J., JMO,
45, 6, 1231-1237, 1998

\bibitem[Arlt et al. (1999)]{arl99}
Arlt, J. and Dholakia, K. and Allen, L. and Padgett, M. J., PR,
59, 5, 3950-3952, 1999

\bibitem[Mair al.(2001)]{mai01}
Mair, A. and Vaziri, A. and Weihs, G. and Zeilinger, A., Nature,
412, 313-316, 2001

\bibitem[Vaziri et al.(2002)]{vaz02}
Vaziri, A. and Weihs, G. and Zeilinger, A., JOB, 4, S47-S51, 2002

\bibitem[Vasnetsov et al.(2005)]{vat}
Vasnetsov, M. V., Pas'ko, V. A., and Soskin, M. S., NJP, 7, 46,
2005

\end{thebibliography}

\end{document}